# *DEMIS*: A Threat Model for Selectively Encrypted Visual Surveillance Data

Ifeoluwapo Aribilola[a,*] (Researcher), Mamoona Naveed Asghar[b,*] (Co-ordinator) and Brian Lee[a] (Co-ordinator)

[a]*Software Research Institute, Technological University of the Shannon: Midlands Midwest, Athlone, University road, Athlone, Westmeath N37 HD68, Ireland*
[b]*School of Computer Science, College of Science and Engineering, National University of Ireland, Galway, Galway H91 TK33, Ireland*

ABSTRACT

The monitoring of individuals/objects has become increasingly possible in recent years due to the convenience of integrated cameras in many devices. Due to the important moments or activities of people captured by these devices, it has made it a great asset for attackers to launch attacks against by exploiting the weaknesses in these devices. Different studies proposed naïve/selective encryption of the captured visual data for safety but despite the encryption, an attacker can still access or manipulate such data. This paper proposed a novel threat model, *DEMIS* which helps analyse the threats against such encrypted videos. The paper also examines the attack vectors that can be used for threats and the mitigation that will reduce or prevent the attack. For experiments, firstly the data set is generated by applying selective encryption on the Regions-of-interests (ROI) of the tested videos using the image segmentation technique and Chacha20 cipher. Secondly, different types of attacks, such as inverse, lowercase, uppercase, random insertion, and malleability attacks were simulated in experiments to show the effects of the attacks, the risk matrix, and the severity of these attacks. Our developed data set with the original, selective encrypted, and attacked videos are available on git-repository for future researchers.

## 1. Introduction

The advancement in technology has introduced different devices with camera integration that protect and monitor the activities carried out in diverse domains. Surveillance devices can perform a recording either in a fixed location (Closed-circuit television (CCTV)) or while in motion (Pan-Tilt-Zoom (PTZ), Dashboard-cam, Unmanned Aerial Vehicle (UAV)). The videos captured from these camera devices mostly contain sensitive and confidential information which makes it an asset that the attackers may exploit by finding vulnerabilities or weaknesses in these devices. After spotting these vulnerabilities, an attack is therefore initiated against the device for the attackers' selfish interest. The encryption of the video was proposed by different studies as a way to protect against these attacks [1, 2, 3, 4, 5]. Despite the videos being sufficiently protected through encryption, an attacker can still exploit security flaws in them.

A threat is a potential risk that exploits a vulnerability (security loophole) to infringe security by causing the confidentiality, integrity, and availability of the information stored or transmitted by the device to be damaged or compromised [6] while an attack is a deliberate act of taking advantage of the weakness in a device [7]. Threats are mainly divided into either a physical threat or a logical threat [6].

The physical threat is a damage of hardware, theft, natural disaster (including flood, fire, war, earthquakes etc) of a system which causes critical data to be stolen or destroyed from the hardware (hard drives or systems) while the logical threat is a vandalisation on the software or data by corrupting the data or exploiting the errors in the software [6].

The impacts of these threats on the daily bases drives the need for a security policy that will be used to safeguard sensitive and confidential information from the attackers. There are different policies like disaster recovery plan (DRP), access control policy (ACP), log management policy (LMP), server security policy (SSP), vulnerability assessment, threat modelling, which can be implemented to protect the system against attacks. Table 1, describes the different terminologies used in this study.

A threat model is a process for ensuring applications and systems security, by identifying objectives, asset, vulnerabilities or absence of appropriate safeguards, and countermeasures to prevent or mitigate these threats. There are different threat modelling techniques which are designed based on the vulnerabilities of the system such as STRIDE [8, 9], attack trees [10], process for attack simulation and threat analysis (PASTA) [11], attack defense trees [12], common vulnerability scoring system (CVSS) [13], security cards [14]. Some threat model techniques like hybrid threat modeling method (hTMM) [15], quantitative threat modeling method (Quantitative TMM) [16] combines two or more of these techniques together. The study [17] provided a good description of these threat models with the threat models evaluating the threats on the network and web, and not on encrypted visual data.

---

[*]This work is part of doctoral degree research funded by the Presidential Doctoral Scheme (PDS) of the Technological University of the Shannon: Midlands, Midwest, Athlone Campus, Athlone, Ireland.
[*]Corresponding Author
[**]Principal corresponding author
i.aribilola@research.ait.ie (I. Aribilola); mamoona.asghar@nuigalway.ie (M.N. Asghar); blee@ait.ie (B. Lee)
ORCID(s): 0000-0003-4488-2925 (I. Aribilola)

**Table 1**
Terminologies.

| Terms | Description |
| --- | --- |
| Surveillance cameras | The cameras that captured and store the activities in a location |
| Visual data/ Surveillance videos | The recorded videos retrieved from the surveillance cameras |
| Asset | An asset is something (here visual data) that is valued, and when compromised can cause a great havoc |
| Vulnerabilities | The security weaknesses or flaws that expose an asset to danger |
| Threats | The potential security risks that may exploit vulnerabilities in assets |
| Attack | An unauthorized attempt or access to alter, remove, acquire, destroy, embed or expose sensitive information in an asset |
| Attack vectors | The ways an attacker uses to gain access to an asset |
| Mitigations | The techniques to prevent threats and attacks |
| Threat model | A structured method that spot potential threats or the absence of appropriate safeguards, quantifies threats and vulnerabilities, and prioritize countermeasures |
| Selective Encryption | A reversible encoding (protection) of data using encryption algorithms (cipher) on specific parts of videos (known as Regions of interest (ROI)) |
| Foreground | The moving objects in the surveillance videos. |
| Background | The static objects in the surveillance videos. |

Different studies has recognised various attacks, that are launched at surveillance videos either in storage or in transmission. The study [18] identified two types of attack against the network of drones which are: intrusion from the outside and network usage from inside. Also the authors [19] proposed a method that protects UAV data from stalkers, plaintext and ciphertext attack using blockchain. The research [20] proposed an application that secures videos from dynamic cameras by storing the foreground objects and the background objects separately to prevent replay attack when the video is in transmission. Threat modelling will help to earlier detect potential threats to guide against unforeseen attacks on these videos.

With numerous attack arising from video retrieved from surveillance cameras and various investigation on threat modelling and attack simulations literature, the following research questions (RQs) identifies the basis of this research:

**RQ1:** What type of attacks produce high impact on protected visual data (asset) retrieved from surveillance camera devices?

**RQ2:** How do the threat model work against the threats on visual data?

**RQ3:** What are the countermeasures and ways to mitigate these threats?

In response to the RQs above, this paper proposed *DEMIS*: a threat modelling technique that protects the most valuable asset (visual data) to prevent potential threats that exploit the security loop-holes within encrypted videos obtained from surveillance camera devices. The research contributions are as follows:

- To expose new and unintended threats by identifying the security flaws, and mitigating the attack against the selectively encrypted videos to improve security postures.

- To ensure the necessary protections are in place and are able to address the evolving threats across the encrypted videos.

- Evaluation of the proposed *DEMIS* threat model against initiated attacks on the selectively encrypted videos.

The remaining parts of the paper are organized as follows: Section II presents related studies on a wide range of attacks and patterns of threat modelling for these attacks. Section III discusses an attack scenario, the methodological analysis behind the proposed *DEMIS* threat modelling application as well as the attack defense tree for encrypted videos. Section IV demonstrates the attack methods against encrypted videos, the statistical analysis of these attacks and also describes how *DEMIS* can be used to mitigate the attacks. The conclusion and future work of this study are presented in Section V.

## 2. Related Work

This section analyses the existing literature on the attacks carried out on the network, web, and surveillance camera devices with their respective threat models. The research gaps are also investigated through this section.

### 2.1. Attacks on the Network security

Network security attacks are unauthorized actions directed at computer information systems, computer networks, infrastructure, or personal computers that are connected together for the purpose of exchanging resources.

The network security attack can be classified into two groups; the passive attack (attackers access, monitor, and steal sensitive information without altering it) and the active attack (attackers damage, encrypt, or modify sensitive information) [21].

Eavesdropping attack, traffic analysis attack, monitoring attack [22, 23], collusion attack, latency attack, message coding attack [24, 25] are examples of passive attacks. The study [26] identifies passive attacks using the key loggers concept and registry files in the system while [25] analysed passive attack in anonymous communication system.

Examples of active attacks are masquerade attack, spoofing attack, Denial-of-Service attack, Wormhole attack [23], fabrication [22] Man-in-the-Middle attack (replay attack) [27], rushing attack, modification attack, node replacement attack [28]. The study [28] describes active attacks on wireless sensor networks (WSNs) with their countermeasures.

#### 2.1.1. Threat models for Network security

For the purpose of guiding against the network security attacks, different threat models like TRIKE, process for attack simulation and threat analysis (PASTA); operationally critical threat, asset, and vulnerability evaluation (OCTAVE); damage potential, reproducibility, exploitability, affected users, and discoverability (DREAD) [29]; visual, agile, and simple threat (VAST), etc are been designed. This section will discuss the STRIDE, attack trees, attack defense trees, and common vulnerability scoring system (CVSS) threat models.

**STRIDE:** STRIDE threat modelling process was introduced by [30] and embraced by Microsoft in 2002 [8] and applied [9]. STRIDE is an acronym for the common threats on the network which are:

*Spoofing:* This attack targets authentication identity, whereby the attacker pretends to be the user (impersonation) to access a system or application.

*Tampering:* This is the modification of data or assert without approval or permission.

*Repudiation:* This is claiming not to be responsible for an action which cannot be verified.

*Information disclosure:* This exposes sensitive data or information to unauthorised persons

*Denial of service:* This is the degradation or exhaustion of a service to the point where users are unable to utilize it normally.

*Elevation of privilege:* This is granting access to a subject that is beyond the subject's authorization permits.

**Attack Trees:** An attack tree is a visual representation that illustrates how an asset might be attacked [31]. In this diagram, the root node, parent node and child node. The root node is always the asset, while child nodes are conditions that must be satisfied for a direct parent node to be true. However, only a node's direct children can be satisfied and once the root node is satisfied, the attack is completed. Attack tree is usually implemented alone, but recently it has been combined with other threat models and frameworks [10]. The study [32] used attack-tree for risk assessment approach.

**Attack Defense Trees (ADT):** ADTs are simple graphical representations of security models and assessment [33, 34]. It is similar to the attack tree but instead of having only the attacks, the defense methods are also included on the tree. The attacks are mostly represented with red circle while the defense method are represented with a green box [12, 35].

**Common vulnerability scoring system (CVSS):** CVSS provides a quantitative severity score for vulnerabilities after capturing their principal characteristics. National Institute of Standards and Technology (NIST) designed CVSS [13] and it is maintained by FIRST (Forum of Incident Response and Security Teams). With the CVSS, users have access to a common and standardized scoring system across a variety of cyber [36] and cyber-physical platforms. CVSS consists of three metrics group which are Base, Temporal, and Environmental metrics [17, 37].

### 2.2. Attacks on the Web security

Web applications are software programs that run on a web server and are accessed by a user via a web browser with an active Internet connection. A user's desire to access the web application triggers a request to the server, which initiates communication with the database. As a result, a response is returned to the web application.

An attack on a web application can take several forms like injection [38, 39], cross-site scripting (XSS) [40, 41], cross site request forgery (CSRF) [42, 43], path traversal [44, 45], Phishing [46, 47], brute-force [48], and distributed denial of service (DDoS) (hTTP flood attacks, session initiation protocol (SIP) flood attacks) [49] attacks.

### 2.2.1. Threat models for Web security

Due to high increase of attacks on the web application, the following techniques are implemented on the server-side and on the client-side as threat model to countermeasure the attacks.

***Server-side Protection:*** DIGLOSSIA, a tool that detects code injection attacks precisely and efficiently on server-side web applications generating SQL and NoSQL queries was proposed by [50]. It is possible to protect the server-side from XSS attack by only allowing "GET" requests to retrieve data, rather than modifying data on the server, and by mandating that all "POST" requests include a pseudo-random value [43]. In order to prevent the CSRF attack, filtering, escaping, sanitizing, validating, and using content security policy (CSP) on untrusted data are all necessary measures[41]. The use of canonical path names, which do not contain tricky references to parent directories or unresolved links, prevents traversal attacks [51]. To reduce the risk of phishing attacks on the server, human exposure should be reduced, and attacks should be detected at the network level or at the end-user device once they are launched [47]. Using specific techniques such as registering new domain names that match well-known domain names can also be used to track down the source of an attack [47]. Among the necessary steps for preventing DDoS attacks are traffic monitoring and analysis, prevention, detection, traceback, characterization, and mitigation [52].

***Client-side Protection:*** This is achieved by creating a tool that acts as a proxy between the client and server, intercepting every HTTP request and deciding whether to allow it or not [43]. For phishing attacks to be prevented on the client-side, end-users need to be educated on how to spot them and avoid taking the bait as well as the use of law enforcement as a deterrent [47]. Long alphanumeric password by using Markov Chain (Markov Password) can resist brute-force attack [53].

### 2.3. Attacks on the Surveillance camera encrypted videos

Surveillance cameras devices like UAV, drones, IoTs are used to observe activities either at a fixed location or on a moving basis. Due to the sensitive nature of the information retrieved from these cameras as videos makes it delicate and attractive to attacker for his or her selfish gain hence, increasing the attack against these videos.

Several attacks has been linked to the encrypted videos from the surveillance camera devices such as intrusion from the outside and network usage from inside [18], data stalking, plaintext and ciphertext attack [19], replay attack [54, 20], exfiltration attack [55], brute-force, known-plaintext and chosen-plaintext attacks [56].

### 2.3.1. Threat models for Surveillance cameras

Despite the awareness of attacks on surveillance camera encrypted videos, many of these studies did not design a threat model to mitigate the attacks. The study [54] proposed a solution to replay attack on Controller Area Network (CAN) bus but did not give a detail of how to mitigate this attack. The study [57] discussed the security analysis of digital image watermarking, but failed to discuss the threat model associated with the technique. Also, [2] designed a lightweight encryption using hybrid chaotic systems but did not analyse the threat to this system.

Several existing research studies reviewed in this study indicate that encrypted video content from surveillance camera devices can be subtly hacked. Therefore, it is extremely important to protect these encrypted videos from attacks. It is also worth pointing out that the existing literature however, does not focus on threat modelling to uncover potential threats on encrypted videos, which constitutes a research gap. On the other hand, this paper proposes the *DEMIS* technique, which reveals the security flaws of encrypted videos from surveillance cameras.

## 3. Proposed Technique

This section analyses the attack-defense tree as well as the proposed threat model technique for encrypted videos from surveillance camera devices.

The asset of the surveillance camera device is the video containing the sensitive information, to safeguard this asset, the encryption of the images in the video is of high priority, but there are loopholes to guard against. Hence, the asset in this study is the encrypted video files.

In this study, the objects in the video was segmented into the foreground (FG) and the background (BG) using advance flow of motion (AFOM) [58] and gaussian mixture method (GMM) [59]. Each of the object pixels was encrypted with Chacha20 cipher algorithm by randomly generating a key and nonce to XOR the pixels, and separately storing the FG from the BG [20].

Chacha20 is a stream cipher algorithm with a complete round of 10 column-rounds and 10 diagonal-rounds while a single round contains four quarter-rounds for a column-round and four quarter-rounds for a diagonal-round [60].

Attack was later simulated against these objects as discussed in the section 4 of this paper. The attack defense tree and the threat model method allow easy identification of security loop-holes on these encrypted videos.

### 3.1. Attack Scenarios on Encrypted Videos

The flowchart in Fig. 1 illustrates an attack scenario for encrypted videos, hence providing a better understanding. There are two ways in which encrypted videos can be attacked: while they are being stored or while they are being transmitted.

An attacker can perform either an injection (adding or extending) of byte to the encrypted frame or can perform a modification of byte to the encrypted frame. After decryption of this video frame, if the attacked frame can be easily detected, then such attack is unsuccessful but if the attacked frame could not be detected, then the attack is a successful one.

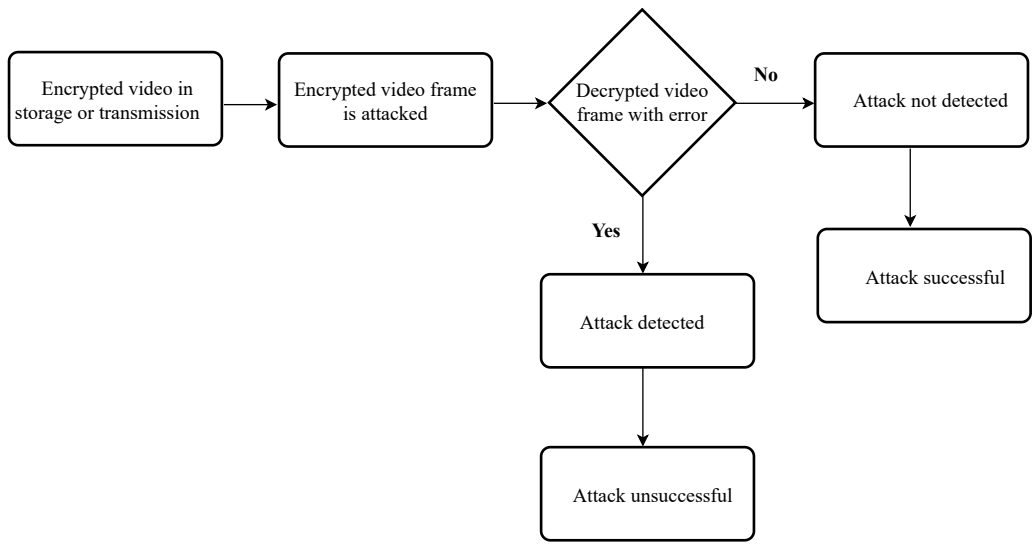

**Figure 1**: Attack Scenario flowchart.

### 3.2. Attack-defense Tree for Encrypted Videos

The ADTs are trees with labeled nodes that are divided into two categories: (1) the Attack nodes (attack) which is the attack that an attacker launches against a system component and (2) the Defense nodes (countermeasures) which are the steps a defender employ to safeguard the system.

As shown in the Fig. 2, the attack nodes are depicted by red circles while the defense nodes are illustrated by green rounded rectangles. Edges represent causal relationships and are disjunctive by default, that is, if one of the conditions of the children node is satisfied, the attack on the parent node is counted as successful. To have conjunctive edges, every nodes condition in the group must be satisfied for the parents node attack to be successful. However, the conjunctive edges can be marked together with connecting arcs by grouping the edges adjacent to a parent node. The broken lines in ADTs represent the countermeasures while the unbroken lines depict the attacks. The arc is the AND conjunction meaning the two conditions needs to be satisfied.

The difference between the attack tree and the ADTs is that the ADTs contains the countermeasures to the attacks but the attack tree only shows the attack or threats to the system.

The ADT design in this study, can be implemented for encrypted videos in storage as well as in transmission. The two main attacks on the encrypted videos can be either an infrastructure attack or a logical attack. The infrastructure attack could be either a network attack, or a natural disaster or an insider disclosing information. The remedy for network attack is using standardized devices which are National Institute of Standards and Technology (NIST) compliance and performing a secured backup. Likewise the prevention of the natural disaster is keeping updates on the disaster forecast and also having a secure backup. For the prevention of the insider attack, training all the staffs especially the newbies not to fall victim of disclosing sensitive information as well as monitoring the staffs. Despite the fact that secure backup can be a countermeasure, it can also be prone to a ransomware attack. To prevent the ransomware, using immutable storage devices or restricting access on storage devices or storing in the cloud with the cloud provider can be a good measure to prevent ransomware. Also, implementing a multi-factor authentication on these previous measures will give a stronger security.

The logical attack on the ADT design as shown in Fig. 2 is the Man-in-the-Middle attack or a replay attack. The countermeasure for this attack is storing the encrypted foreground object separately from the encrypted background objects as experimented in [20]. The attacker will not be able to accurately detect where to slot in the replayed video. Even with this countermeasure, an attacker can perform a byte injection or byte modification on the encrypted video. The byte injection can be either a random byte attack or a malleability attack (Extending the bytes). The byte modification can be an inverse replacement attack, a lowercase replacement attack or an uppercase replacement attack. These attacks can be prevented by daily running checks on the network to detect spoofing or eavesdropping alongside storing the encrypted video as a Write Once Read Many (WORM) to avoid modifying or adding to the byte. Also using a multi-factor authentication on the WORM stored video will give a stronger protection.

### 3.3. *DEMIS* Threat Model technique

A threat model identifies and characterizes potential threats to a system in an orderly manner. As part of the complete assessment of the threat model analysis, it prioritizes threats and mitigation based on probabilities, impacts, and countermeasure costs. It evaluates all possible risks throughout the system rather than focusing only on the parts of the system that are expected to exhibit flaws.

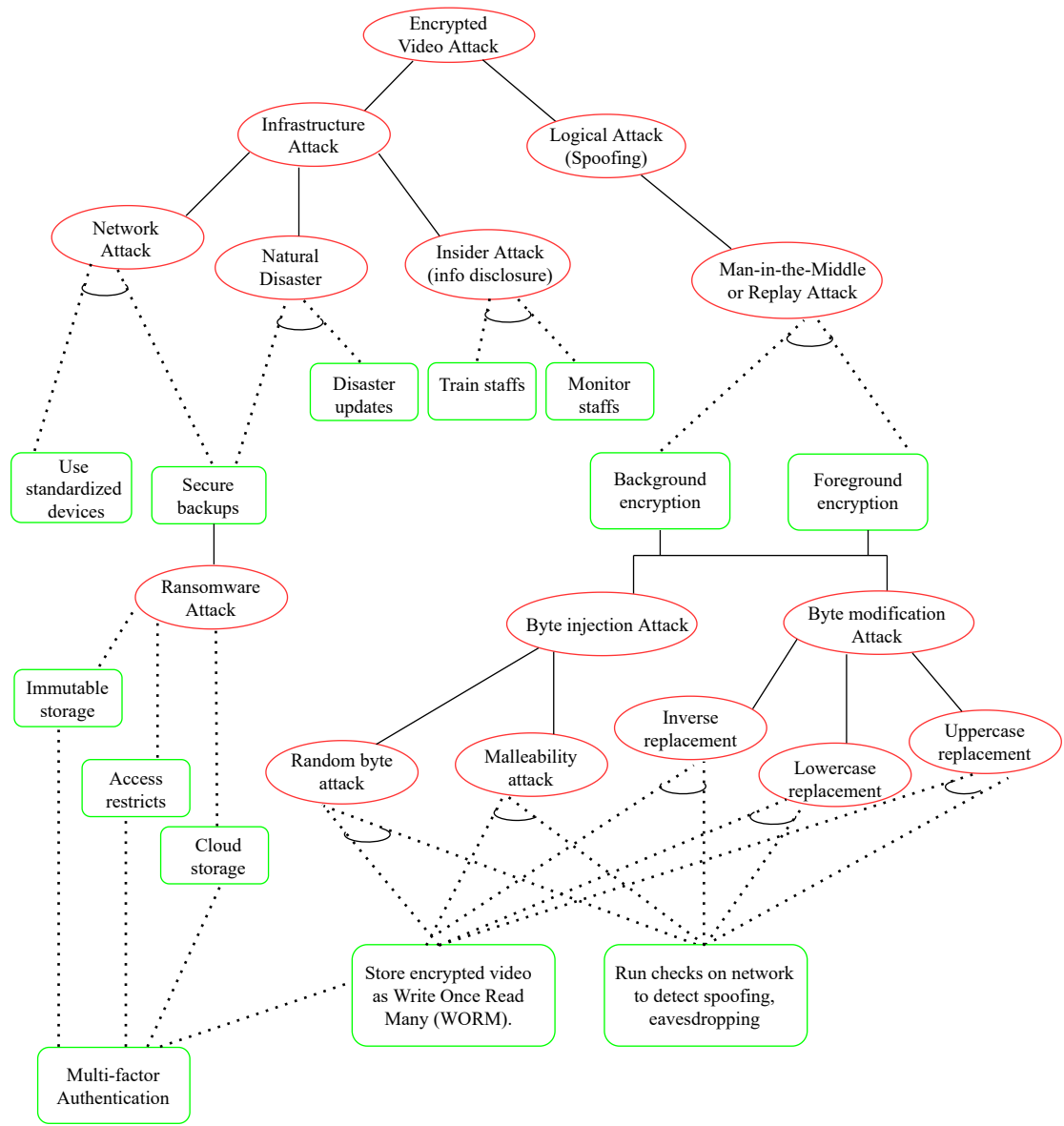

**Figure 2:** Attack Defense tree.

The *DEMIS* threat model technique, is an acronym for the possible threats that can be aimed against encrypted videos, including their possibility, impact, and mitigation, either while the video is in transit or during storage as displayed in Fig. 3. In Fig. 3, the attack vectors are the ways the attacks can be instigated. The vulnerabilities are system flaws, while the mitigation are preventive measures. The asset in this situation is the encrypted video (that is the target of the attacker) while the threat to this asset is the acronym *DEMIS* and is further explained as follows:

***Defects on the network:*** The encrypted videos are transmitted over the network, thus a defect on the network can be disastrous to these videos and can cause these encrypted videos to be accessed by unauthorised persons by spoofing the IP (internet protocol) addresses, session hijacking, which is of a high risk. This attack can be achieved if there are wrong configurations and installation of security policies on network devices like not updating the security patches of the operating system on the computers. Preventing this attack requires using standardized devices for network configuration and installation, updating all network devices as well as the security patches, and performing secure backups to avoid unforeseen future circumstances.

***Exposure of Information:*** Information exposure is a high risk of attacking encrypted video files. While the encrypted video files are in storage, an experienced dissatisfied or a newbie employee can be lured or manipulated to disclose confidential information (password or keys) to access the

**DEMIS: Threat model for encrypted videos**

| Asset | Threats | Attack vectors | Vulnerabilities | Mitigations |
|---|---|---|---|---|
| Encrypted video | Defects on the network | Distributed Denial-of-Service (DDoS), IP spoofing, session hijacking | Wrong configuration and installation of security policies on network devices | Secured backups, use standardized and updated devices |
| | Exposure of Information | Experienced dissatisfied employee, a newbie employee | Lack of user privilege access, weak or common password policies, disclosing confidential information | Training and monitoring of staffs, giving staffs least privilege access |
| | Modification of bytes | Performing inverse attack, lowercase attack, uppercase attack | Password spying, using common or weak password, illegal accessing, broken authentication, insecure encryption storage | Store encrypted video as Write Once Read Many (WORM), using least privilege access, multi-factor authentication |
| | Injection of bytes | Executing random byte insertion attack, extending the byte attack | Password spying, using common or weak password, illegal accessing, broken authentication, insecure encryption storage | Store encrypted video as Write Once Read Many (WORM), using least privilege access, multi-factor authentication |
| | Spoofing of the video content | Eavesdropping, replay attack, content spoofing, content hijacking | Unsecured network, lack of firewall, video not segmented into foreground and background | Storing foreground encrypted objects separately from the background encrypted objects |

**Figure 3:** *DEMIS* Threat modelling technique for attack against encrypted videos.

system where the files are stored. Training and monitoring of staff as well as granting staff the least privileged access to the system will be a corrective measure to avoid this type of attack.

*Modification of Bytes:* Modification of the encrypted bytes by an attacker is a high risk of attack. This is an illegal accessing of the encrypted bytes to either twist the file content to benefit the attacker or render the video useless or inaccessible by altering it, thus the purpose of the video is nullified. The modification attack can be a lowercase attack, an uppercase attack, or an inverse attack (these attacks are further explained in section 4 of this study). It can affect files stored when the password to the storage is weak, spied on, authentication is broken or faulty, or the storage is left open by mistake. To prevent this attack it is advisable to store encrypted video files as a Write Once Read Many (WORM), implement least privilege access, and use multi-factor authentication.

*Injection of Bytes:* The injection of encrypted bytes is similar to the modification of bytes attack but instead of modifying the bytes, the attacker adds or extends the encrypted bytes with his or her own bytes by implementing a random byte attack or a malleability attack (extending the encrypted bytes) making the attack of high risk. This is an illegal falsification of the stored video content and can be accomplished when there is a broken authentication, weak password usage, spied password, or illegal accessing of the system. It is recommended that encrypted video files are stored as Write Once Read Many (WORM), least privilege access be implemented, and multi-factor authentication be used in order to prevent this attack.

*Spoofing of the Video content:* Spoofing of the encrypted video can be executed during transmission of the video files with eavesdropping, Man-in-the-Middle, or replay attacks. An attacker can maximize this attack by introducing false or old content into the video hence this attack is of high risk. The attacker can carry out this attack by exploiting the flaws of an unsecured network, lack of firewalls, or unsegmented encrypted videos in the foreground and background. Separately, storing the encrypted video's foreground from the background will prevent the attacker from accurately introducing the false video content.

## 4. Experimental Results and Analysis

This section discussed the attack simulation, risk analysis and the severity impacts of these attacks on the foreground encrypted video files.

**Table 2**
Properties of the datasets used for the experiments.

| File | Background | Resolution | Rate | Count |
| --- | --- | --- | --- | --- |
| Horse moving | dynamic | 860 × 484 | 23 | 126 |
| MOT16-12 | dynamic | 960 × 540 | 30 | 180 |
| Car moving | dynamic | 1280 × 720 | 30 | 180 |
| Highway | static | 1280 × 720 | 25 | 127 |
| PET | static | 768 × 576 | 7 | 84 |
| Mall | static | 960 × 540 | 25 | 200 |

The simulation of the attacks launched against the encrypted videos was implemented in python programming language and the experiment was performed on the system specifications of Intel(R) Core (TM) i7-10510U CPU @ 1.80GHz 2.30GHz processor, 16GB RAM with an Intel(R) UHD graphics.

The experiment was conducted using a dataset of three (03) dynamic background and three (03) static background giving a total of six (06) publicly available surveillance camera videos [61, 62, 63] captured from dynamic and static camera devices with different characteristics, including colour, motion activity, and spatial details. The test video properties are described in Table 2, where the frame rate per seconds (FPS) represents the frame rate of the videos.

The test video files were segmented into the foreground (FG) and background (BG) using the GMM [59] for the static camera videos and AFOM [58] for the dynamic camera videos; these segmented video files were encrypted using Chacha20 algorithm [60] as analysed in [20]. The FG are the objects in motion while the BG are the static objects in the video.

A single frame was selected from each of the FG encrypted video files and different attacks was launched at it. Table 3, analysed the selected video frames by showing the original frame on row 1, the FG encrypted frames with no attack on row 2, and the decrypted frame with no attack on row 3.

The attacks were also simulated on some other frames of the FG encrypted video files and these are stored on our github repository [64] as well as the FG encrypted with attack frame and decrypted with attack frame for future access by other researchers.

## 4.1. Simulation of Attacks

This research simulated attacks against the FG encrypted video files (asset) to describe how these attacks can damaged these files after decryption.

The byte modification attack and the byte injection attack are further explained with their examples.

### 4.1.1. Simulation of Byte Modification Attack

The byte modification attack is an alteration of the encrypted bytes which can be divided into three (03) forms. The byte modification attack was instigated against the selected FG encrypted frames in Table 3 (b1 - b6), and the attacked results on the the decrypted files are shown in Fig. 4.

**The Inverse Replacement Attack:** This attack was executed on the encrypted FG segment of test videos from both static and dynamic camera devices by replacing all the zeros(0's) digit in the encrypted bytes to ones (1's) digit. The result of the inverse replacement attack is shown in Fig. 4(a1 - a6).

Comparing the decrypted without attack frame in Table 3(c1 - c6) and the decrypted with attack in Fig. 4(a1 - a6), the decrypted frame with attack resembles the respective decrypted frame without attack. This indicates that little or no modification effect is clearly seen and it could also be interpreted as the encrypted bytes has fewer zeros (0's) digits.

In conclusion, an inverse attack of changing zeros (0's) digit to ones (1's) digits in an encrypted video can not be visibly detected. Therefore, this makes the attack very dangerous to the encrypted video files.

**The Lowercase Replacement Attack:** The lowercase replacement attack was achieved on the segmented FG of the six (06) encrypted test video files by converting all the alphabetical letters in the encrypted bytes to their corresponding lowercase letter. The result of the single frame selected from the test videos after the attack are displayed in Fig. 4 (b1 - b6).

This lowercase replacement attack is very effective on the video frames preventing a clear visibility of the FG objects in the video with a black coloration. However, the lowercase attack was not effective on the highway video file in Fig. 4(b4).

The no effect of the lowercase replacement attack on Fig. 4(b4) could mean few lowercase alphabetical letters are contained in the encryption bytes of the particular frame.

An attacker can implement the lowercase attack to cause a large distortion on the video after decryption thus rending the video useless or ineffective.

**The Uppercase Replacement Attack:** In order to perform the uppercase replacement attack, all alphabetical letters in the encrypted bytes were converted to their uppercase equivalents in the segmented FG of the six (06) encrypted test video files as shown in the selected video files in Fig. 4(c1 - c6). In Fig. 4(c4 and c5), the FG objects can be identified, thus the uppercase attack could break through some of the encrypted test video files, posing as a threat to the asset.

Also, this attack affected all the test video frames which conveys its sensitivity on encrypted video files.

**Table 3**
Visual representation of the original frame, encrypted frame and decrypted frames before attack simulation

| Horse moving | MOT16-12 | Car moving | Highway | Mall | PET |
| --- | --- | --- | --- | --- | --- |
| 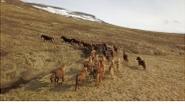 | 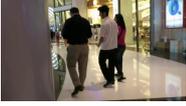 | 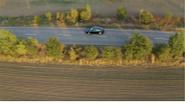 | 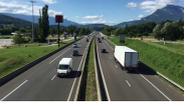 | 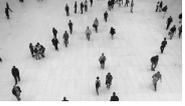 | 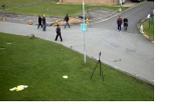 |
| a(1) (Frame # 40) Original | a(2) (Frame # 60) Original | a(3) (Frame # 50) Original | a(4) (Frame # 60) Original | a(5) (Frame # 60) Original | a(6) (Frame # 40) Original |
| 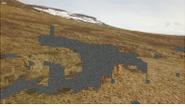 | 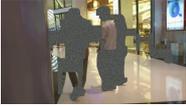 | 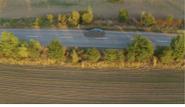 | 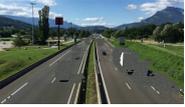 | 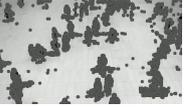 | 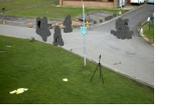 |
| b(1) FG Encrypted with no attack | b(2) FG Encrypted with no attack | b(3) FG Encrypted with no attack | b(4) FG Encrypted with no attack | b(5) FG Encrypted with no attack | b(6) FG Encrypted with no attack |
| 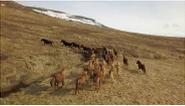 | 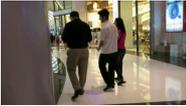 | 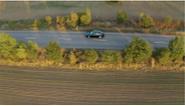 | 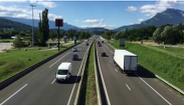 | 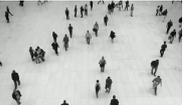 | 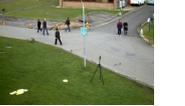 |
| c(1) Decrypted with no attack | c(2) Decrypted with no attack | c(3) Decrypted with no attack | c(4) Decrypted with no attack | c(5) Decrypted with no attack | c(6) Decrypted with no attack |

### 4.1.2. Simulation of Byte Injection Attack

The byte injection attack is an insertion of the bytes into the encrypted video and it is categorised into two (02) forms. The byte injection attack was launched against the the selected FG encrypted frames in Table 3 (b1 - b6) and the effect of these attacks after decryption could be easily spotted as displayed in Fig. 4.

**The Random Byte Insertion Attack:** This attack was executed by injecting random bytes into the FG encrypted bytes of the video files. Eight different random bytes was injected into different position of the encrypted byte video file.

Comparing the decrypted frame without attack in Table 3 (c1 - c6) with the attacked decrypted frame in Fig. 4(d1 - d6), the attacked decrypted frames in Fig. 4(d1 - d6) shows a visible difference with the respective decryption without attack frame which indicates that there were some modifications on the video files.

The random attack in Fig. 4(d1 - d3) shows the visibility content of the video while the visibility of the FG objects in Fig. 4(d4 - d6) can not be identified.

Nevertheless, the random attack is a threat to the encrypted video files (asset).

**The Malleability Attack:** This attack was launched by extending the FG encrypted bytes with a set of bytes. These bytes were added to make the FG encrypted bytes longer.

After decryption, it was observed that the moving objects in the frames could not be properly identified as shown in Fig. 4(e1 - e6) hence, we can conclude that this attack is of a high risk to the encrypted objects in the video.

Finally, all these attacks has a great impacts on the encrypted video after decryption, but the malleability attack has a larger impact than the other attacks. In a situation, where images transmitted by an autonomous vehicle is been encrypted and one of these attacks was launched by an attacker against it, the autonomous vehicle will not be able to get a clear information of the image therefore causing a fatal accident.

### 4.2. Statistical Analysis of the Attack

The statistical analysis of the simulated attacks on the decrypted with attack frames was evaluated using the entropy distribution of the frame's information content, histogram analysis of the pixel colors, the mean square error of the frame and the SSIM values with the original frame.

### 4.2.1. Entropy Analysis

Entropy is a quantitative measure of the information content of an image by quantifying the amount of uncertainty or randomness within the image, therefore the more information content it contains, the higher its quality will be [65]. Hence, the entropy (information content) is directly proportion to the image quality.

The entropy is calculated on a grayscale pixel of 8-bit with a probability ratio of 1:256 in each case, if the information content of a grayscale image is close to 8, it implies that the image has good randomness and quality, the lower the entropy the lower the randomness and quality.

$$H(Y) = -\sum_{i=1}^{x} p(i) \log_2 p(i) \quad (1)$$

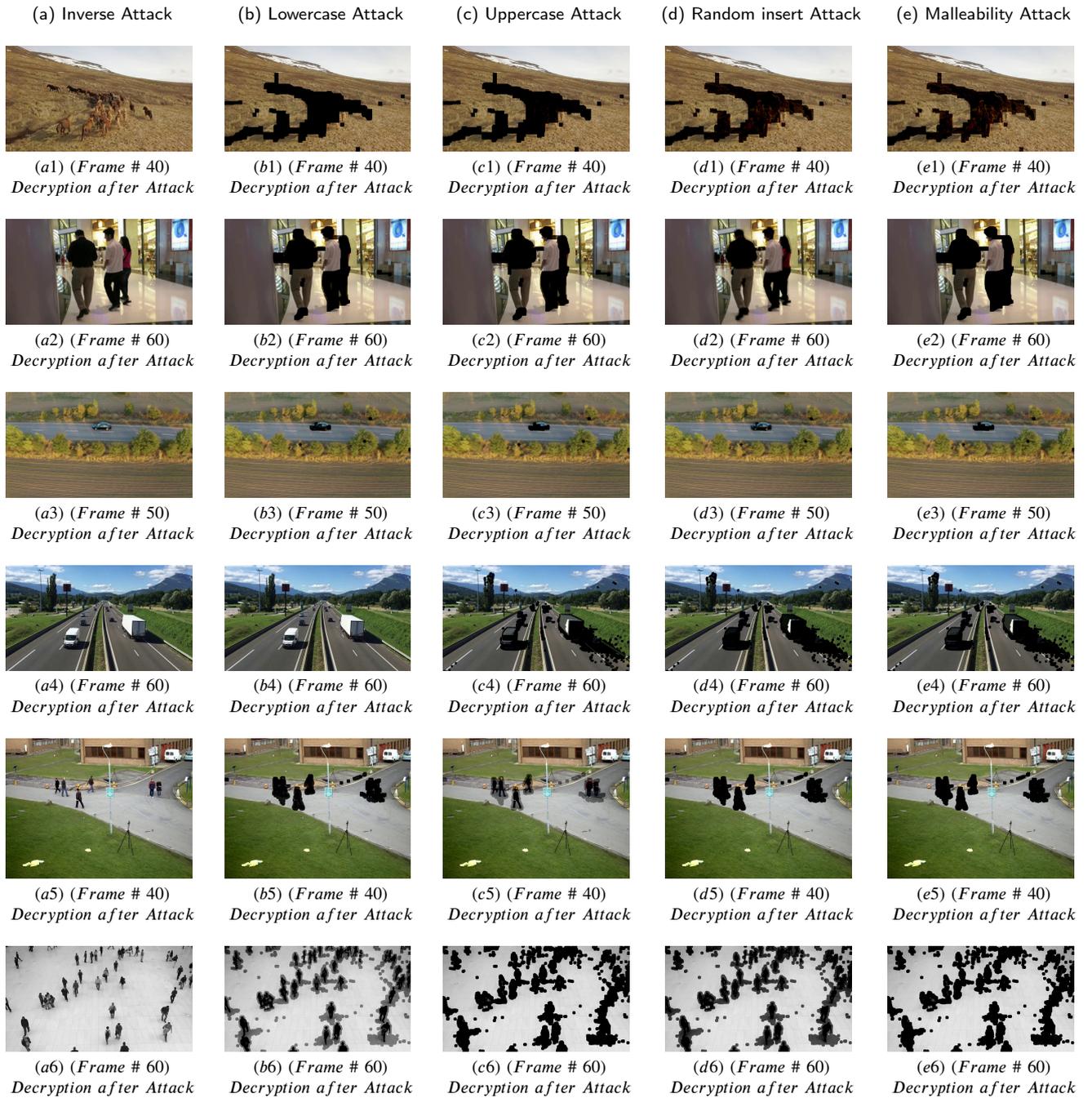

**Figure 4:** Visual representation of decrypted test videos video frames after attack (a1-a6) decryption after inverse attack, (b1-b6) decryption after lowercase attack, (c1-c6) decryption after uppercase attack, (d1-d6) decryption after random insert attack, (e1-e6) decryption after malleability attack.

Entropy was calculated using Equation. 1, where Y is a discrete random variable, x is the number of the gray values distributed from the interval of [0, 255], p(i) is the probability of occurrence of the gray values, and the logarithm base set to 2, return bits value.

The entropy results of the original frame and decrypted with attack frame are discussed in Table 4 for the video files. The entropy value of the original frames are higher than the decrypted with attack frames. The attacked frames with lower entropy are the frames with high visibility of the attack where the FG objects can not be identified while the FG objects in the attacked frame with a closer value to the original frame can be recognised. However, comparing the entropy result of the inverse attack frame with the original frame are equal or in a close value range, which indicates that the inverse attack has no or little impact on the encrypted video files. The entropy reveals that when an attack is launched against the frame, the randomness and the quality of the image decreases.

**Table 4**
The Entropy results of the original frame and decryption with attack.

| Video file | Background | Frames | Original | Inverse attack | Lowercase attack | Uppercase attack | Random attack | Malleability attack |
|---|---|---|---|---|---|---|---|---|
| Horse moving | Dynamic | (*Frame* # 40) | 7.1385 | 7.1385 | 6.5184 | 6.6235 | 7.0447 | 7.0992 |
| MOT16-12 | Dynamic | (*Frame* # 60) | 7.4812 | 7.4812 | 7.0274 | 7.0143 | 7.2180 | 6.9283 |
| Car moving | Dynamic | (*Frame* # 50) | 6.3396 | 6.3396 | 6.3482 | 6.3380 | 6.3562 | 6.3390 |
| Highway | Static | (*Frame* # 60) | 7.4194 | 7.4171 | 7.4171 | 7.2419 | 7.2154 | 7.2338 |
| PET | Static | (*Frame* # 40) | 7.3733 | 7.3725 | 7.3974 | 7.3672 | 7.3114 | 7.3066 |
| Mall | Static | (*Frame* # 60) | 6.5026 | 6.4995 | 6.7785 | 5.1628 | 6.5988 | 5.3744 |

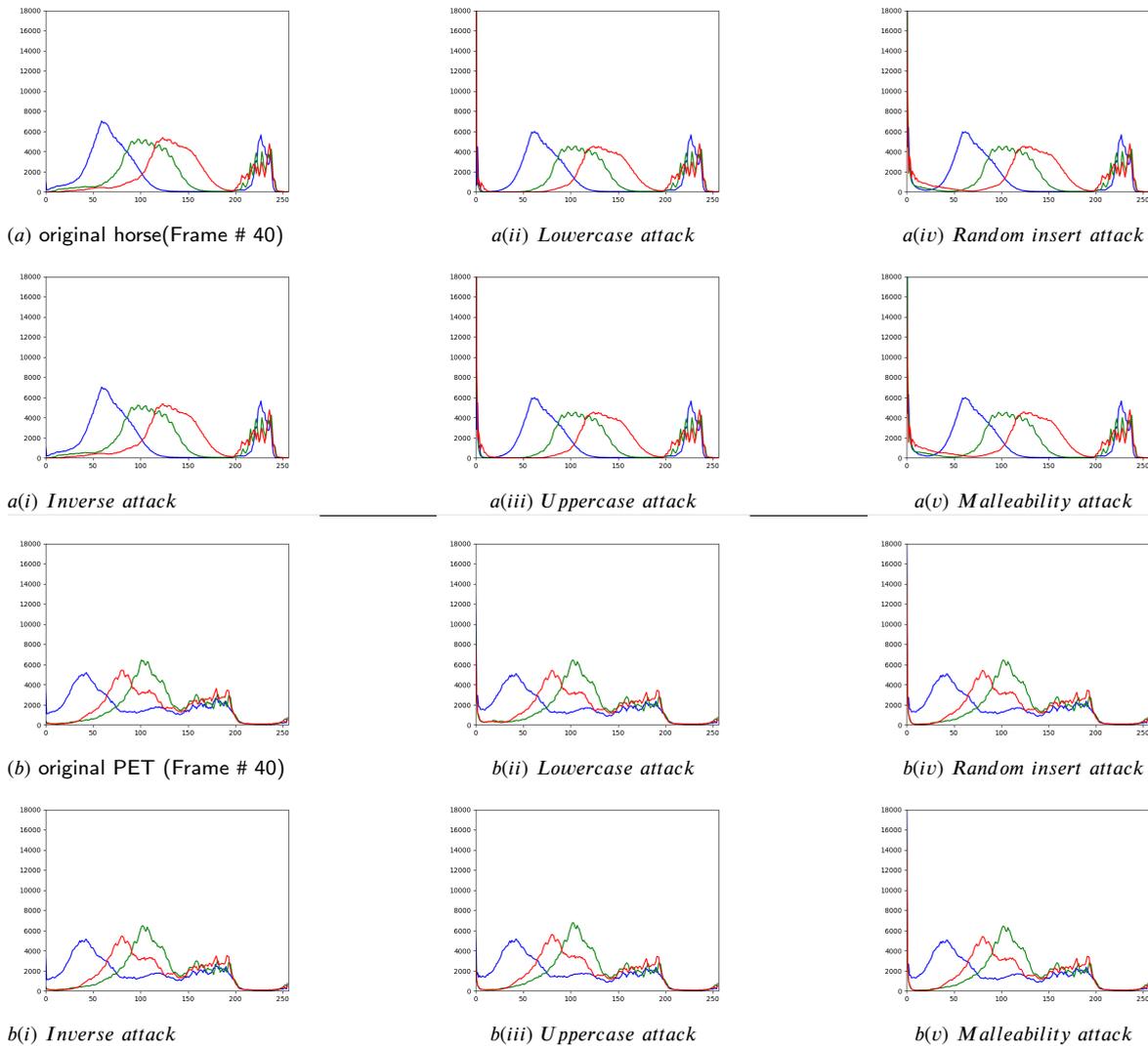

(*a*) original horse(Frame # 40)    *a(ii) Lowercase attack*    *a(iv) Random insert attack*

*a(i) Inverse attack*    *a(iii) Uppercase attack*    *a(v) Malleability attack*

(*b*) original PET (Frame # 40)    *b(ii) Lowercase attack*    *b(iv) Random insert attack*

*b(i) Inverse attack*    *b(iii) Uppercase attack*    *b(v) Malleability attack*

**Figure 5:** Histogram analysis of the simulated attacks performed on test videos (a, b) original video frames, (a(i), b(i)) decrypted video frames after inverse attack, (a(ii), b(ii)) decrypted video frames after lowercase attack, (a(iii), b(iii)) decrypted video frames after uppercase attack, (a(iv), b(iv)) decrypted video frames after random insert attack, (a(vi), b(vi)) decrypted video frames after after malleability attack.

**Table 5**
The Mean-Square Error results of the original frame and decryption with attack.

| Video file | Background | Frames | Inverse attack | Lowercase attack | Uppercase attack | Random attack | Malleability attack |
|---|---|---|---|---|---|---|---|
| Horse moving | Dynamic | (Frame # 40) | 0 | 19.7253 | 19.6788 | 20.6856 | 20.9470 |
| MOT16-12 | Dynamic | (Frame # 60) | 0 | 11.4693 | 11.8454 | 13.5529 | 11.6761 |
| Car moving | Dynamic | (Frame # 50) | 0 | 0.8806 | 0.8593 | 0.8079 | 0.8599 |
| Highway | Static | (Frame # 60) | 7.7469 | 7.7469 | 17.6639 | 17.8957 | 18.0866 |
| PET | Static | (Frame # 40) | 3.7717 | 6.1203 | 7.1237 | 6.4292 | 6.4325 |
| Mall | Static | (Frame # 60) | 8.7282 | 38.2510 | 36.5319 | 38.0119 | 37.7647 |

### 4.2.2. Histogram Analysis

The histogram analysis, is a plot of the frequency distribution of the pixels' intensity values based on the RGB (red, green, and blue) colour component of the original frame, and the decrypted video frames with simulated attack.

In this research, the histogram analysis of the attacks was executed on two video files (a dynamic and a static background) and the result is shown in Fig. 5. The original frame and the decrypted with inverse attack frame in Fig. 5((a), a(i)), have their RGB analysis originating from zero on the y-axis, while the decrypted with lowercase attack, the decrypted with uppercase attack, the decrypted with random insert attack and the decrypted with malleability attack have their RGB histogram analysis beginning from the peak of y-axis.

Also, in Fig. 5((b), b(i)), the original frame and the decrypted with inverse attack frame are similar having their red and green pixels starting from the origin on the y-axis of the graph. However, the red and green pixels of the decrypted with lowercase attack, the decrypted with uppercase attack, the decrypted with random insert attack and the decrypted with malleability attack do not start from the origin as shown in Fig. 5(b(ii), b(iii), b(iv), b(v)).

The histogram analysis of the decrypted with inverse attack demonstrates that the inverse attack can not be detected on the frame unlike the other attacks (lowercase, uppercase, random insert, malleability).

### 4.2.3. Mean-Square Error (MSE) Analysis

The Mean-Square Error (MSE) evaluates and compares the quality of image compression. The MSE measures the cumulative squared error difference between the attacked frame and the original frame, therefore the lower the MSE values, indicates a lower error values. The MSE was calculated using Equation. 2, where n is the number of pixels, $Y_i$ is the original frame and $\hat{Y}_i$ is the attacked frame.

$$MSE = \frac{1}{n} \sum_{i=1}^{n} (Y_i - \hat{Y}_i)^2 \qquad (2)$$

The mse value of the inverse attack column in Table 5 has the lowest result value and this implies that the inverse attack was not effective on these video files. The frames with high impact of the attacks are with high mse values in Table 5.

### 4.2.4. Structural Similarity (SSIM) Analysis

SSIM is measured by comparing the perceptual difference between two images. It performs comparison base on the structure, luminance and contrast of the pixels in the image using a range of -1 to +1. The SSIM was calculated using Equation. 3.

$$SSIM(x, y) = \frac{(2\mu_x\mu_y + c1)(2\sigma_{xy} + c2)}{(\mu_x^2\mu_y^2 + c1)(\sigma_x^2 + \sigma_y^2 + c2)} \qquad (3)$$

where x = original tested videos, y = decryption with attack videos, $\mu_x$ = average of x, $\mu_y$ = average of y, $\sigma_x^2$ = variance of x, $\sigma_y^2$ = variance of y, $\sigma_{xy}$ = covariance of x and y, $c1 = (K_1L)^2$ $c2 = (K_2L)^2$, L = dynamic range, $(K_1) = 0.01$, $(K_2) = 0.03$

The closer the SSIM value to +1, the lesser the variance between the images. Table 6 analysed the results of the comparison between the original frames and the attacked frames, the static background mall test video has the lowest SSIM values among all the test video files.

### 4.3. Security Analysis of *DEMIS* Threat Model

The security analysis of the *DEMIS* threat modelling technique was evaluated using risk assessment matrix, CVSS calculator to check the impact of the threats when the mitigation are applied.

### 4.3.1. Risk Assessment Matrix for the Threats

A risk assessment matrix was designed to analyze the impact of potential threats and their probabilities of occurrence from the perspective of the attack vectors, and in turn it identifies and prevents the high impacted and highly occurring attack vectors so as to minimize the potential for attacks against the asset.

The risk matrix in Fig. 6 shows the impacts and likelihood of occurrence of the attack vectors for each of the identified threats. In Fig. 6, all the attack vectors that can exploit these vulnerabilities in the asset are of high risk, thus, safeguarding these risks using the mitigation stated in Fig. 3 column 5 will help prevent these attacks.

**Table 6**
The SSIM results of the original frame and decryption with attack.

| Video file | Background | Frames | Inverse attack | Lowercase attack | Uppercase attack | Random attack | Malleability attack |
|---|---|---|---|---|---|---|---|
| Horse moving | Dynamic | (*Frame* # 40) | 1 | 0.8028 | 0.8033 | 0.8149 | 0.8188 |
| MOT16-12 | Dynamic | (*Frame* # 60) | 1 | 0.8649 | 0.8641 | 0.9063 | 0.8622 |
| Car moving | Dynamic | (*Frame* # 50) | 1 | 0.9914 | 0.9907 | 0.9939 | 0.9907 |
| Highway | Static | (*Frame* # 60) | 0.9502 | 0.9502 | 0.8533 | 0.8472 | 0.8491 |
| PET | Static | (*Frame* # 40) | 0.9740 | 0.9345 | 0.9496 | 0.9328 | 0.9327 |
| Mall | Static | (*Frame* # 60) | 0.9504 | 0.7198 | 0.6042 | 0.6598 | 0.6049 |

**Figure 6:** Risk assessment matrix for the threats against encrypted videos (asset) showing the likelihood and impacts of the attack vectors.

### 4.3.2. CVSS calculator Analysis

A common vulnerability scoring system (CVSS) calculator managed by FIRST.Org is an open and free framework for communicating the characteristics and severity of vulnerabilities related to computer systems and software. There are three metric groups in CVSS: Base, Temporal, and Environmental; Base group represents the intrinsic characteristics of a vulnerability that are persistence over time and across user environments; the Temporal group reflects vulnerability characteristics as they change over time; and the Environmental group represents vulnerability characteristics specific to a user's environment.

Fig. 7 analysed the base score of the threat's severity using the CVSS calculator, The impact which is the right hand side of the base score was common to all the threats, Scope (S): the impact is only localized to the vulnerable component, Confidentiality (C): if attacked, all the information is disclosed to the attacker, Integrity (I): the attacker can modify the information and, Availability (A): the information could be non available or mutated. Hence, the scope, confidentiality, integrity and availability (SCIA) are set to "High" because a breach will greatly affect the asset.

On the left hand side of Fig. 7, the attack complexity (AC), privilege required (PR) and user interaction (UI) are the same for the threats, except the exposure of information threat. AC, PR and UI are low, none and none respectively, the low AC means attack can exploit the vulnerability anytime, none PR means the attacker is unauthorised while the none UI means the attacker can achieve the attack without user interaction. But for the exposure of information the attacker needs a UI to achieve the attack. The attack vector (AV) depends on if the attack is on the network, through a local network like wifi or Bluetooth, attacker could log in locally, or attacker requires a physical access.

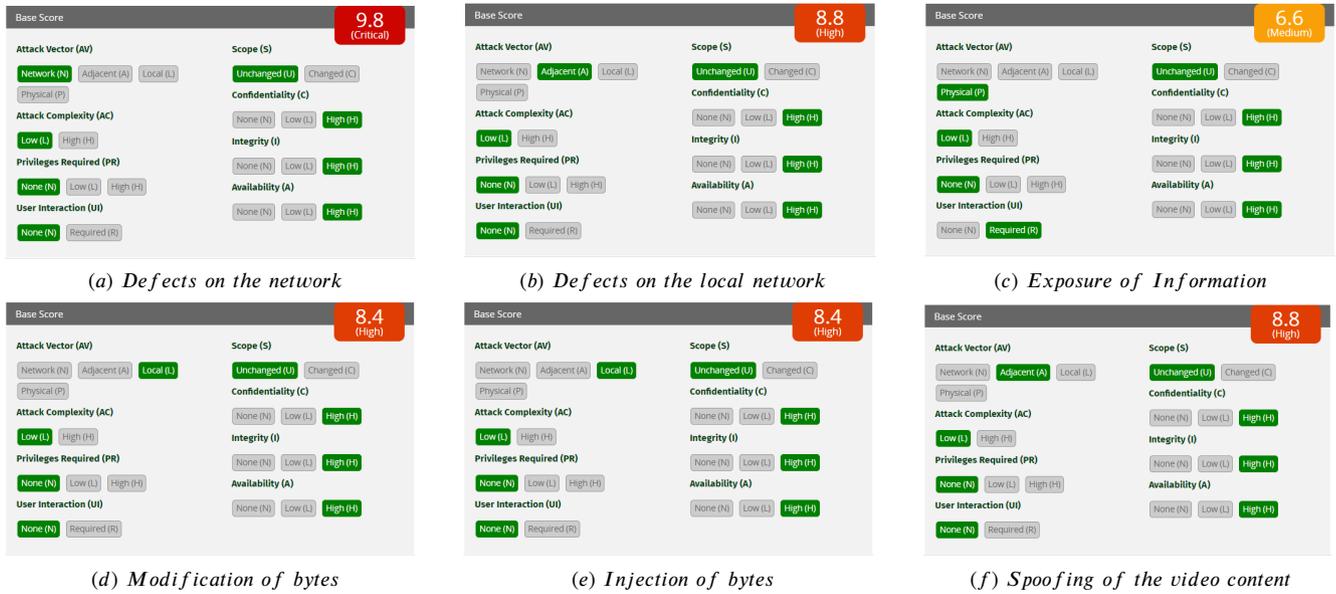

**Figure 7:** Threat severity impact analysis using the common vulnerability scoring system (CVSS) calculator

Implementing the several mitigation analysed in Fig. 3 column 5 will reduces these severity.

***Using Standardized devices:*** Using standardized and security patch updated devices for network configuration and installation will reduce the severity of these attacks.

***Training Staffs and Least Privilege Access:*** Training the staffs will give a newbie and other staffs a clear understanding of the attacker manipulative techniques. Also, monitoring of staffs by giving least privilege access for easy accountability if there are any suspicious moves.

***Multi-factor Authentication:*** This will increase the security of the system from been easily accessed by anyone guessing or spying on the password.

***Using a WORM device:*** A WORM (Write Once Read Many) will prevent an attacker from modifying or injecting into the encrypted video files (asset).

***Storing the FG objects separately from the BG objects:*** Storing the FG separately from the BG objects will prevent an attacker (who only could access one of the files) from correctly identifying the file to manipulate.

### 4.4. Comparative analysis of *DEMIS*

The proposed threat modelling technique *DEMIS* was compared with existing state-of-the-art (SOTA) threat models in Table 7.

From Table 7, it can be deduced that only the proposed *DEMIS* threat model investigated the threats severity impacts, the risks associated with exploiting the threats with the attack vectors and the statistical evaluation of these attacks.

## 5. Conclusion and Future work

Encryption is the only reversible technique for protecting data. For efficiency, selective encryption (on specific ROI) is highly applicable these days for low-powered computation devices. This paper devised a novel threat model *DEMIS* for the selective encrypted videos (valuable asset) captured from the surveillance camera devices. To the best of our knowledge, this study is a pioneer in developing threat model for protected visual data. *DEMIS* analysed different attack simulations against the asset by recognising the threats and the attack vectors to exploit the threats on the encrypted videos. *DEMIS* can easily detect potential threats through the risk assessment impact as well as provide the countermeasures to avert or lessen these threats. *DEMIS* proved to be a productive threat model for encrypted videos captured from surveillance camera devices.

In the future extension of this research work, a model will be designed that will be able to classify the original frame and if a surveillance video frame has been attacked. The classification of these attacks will assist to predict the type of attack that was launched on the video and finally prevent or remove the attacked video frames from the network in order to safeguard the network.

**Table 7**
Comparison of *DEMIS* with SOTA threat models.

| | STRIDE [30] | OCTAVE [66] | VAST [67, 68] | PASTA [11] | Trike [69] | ML SDA [70] | Proposed *DEMIS* |
|---|---|---|---|---|---|---|---|
| Identify suitable mitigating control | ✓ | ✓ | ✓ | ✓ | ✓ | ✓ | ✓ |
| Classify threat mitigation efforts | ✗ | ✓ | ✓ | ✓ | ✓ | ✓ | ✓ |
| Ability to detect new threat | ✗ | ✗ | ✗ | ✗ | ✗ | ✓ | ✓ |
| Ability to identify multiple threats | ✗ | ✗ | ✗ | ✗ | ✗ | ✓ | ✓ |
| Threat severity impacts | ✗ | ✗ | ✗ | ✗ | ✗ | ✗ | ✓ |
| Contribute to risk management | ✗ | ✓ | ✓ | ✓ | ✓ | ✗ | ✓ |
| Analyse the risks of exploiting the threats with the vector attack | ✗ | ✗ | ✗ | ✗ | ✗ | ✗ | ✓ |
| Simulation of attacks | ✗ | ✗ | ✗ | ✓ | ✗ | ✗ | ✓ |
| Statistical analysis of the attacks | ✗ | ✗ | ✗ | ✗ | ✗ | ✗ | ✓ |

## CRediT authorship contribution statement

**Ifeoluwapo Aribilola:** Data curation, Introduction, Literature Review, Methodology, Experiments, Writing - Original draft preparation. **Mamoona Naveed Asghar:** Conceptualization of this study, Methodology, Writing - Original draft preparation
(*Authors work equally for this Manuscript).

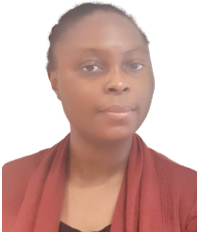

Ifeoluwapo Aribilola received her B.Tech degree from the Department of Computer Science, School of Science, Federal University of Technology Akure, Nigeria, in 2012, her M.Sc. degree from the Department of Computer and Software Engineering, Faculty of Engineering & Informatics, Athlone Institute of Technology Athlone, Ireland, in 2019, and currently pursuing a Ph.D. degree with the Software Research Institute, Technological University of the Shannon: Midlands Midwest, Athlone campus, Athlone, Ireland. She has over seven years of working experience as a software engineer in different IT companies. Her research interests includes and not limited to deep learning, computer vision, visual privacy, threat modelling, cryptology, cybersecurity.

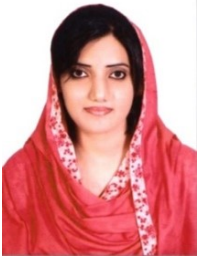

Mamoona Naveed Asghar received a Ph.D. degree from the School of Computer Science and Electronic Engineering, University of Essex, Colchester, U.K., in 2013. She is a former Marie Sklodowska-Curie (MSC) Career-Fit Research Fellow at AIT-Ireland (2018-2021). She is presently working as an Assistant Professor in the School of Computer Science, COSE, National University of Ireleand, Galway, Ireland. She has published several ISI indexed journal articles along with numerous international conference papers. She is actively involved in reviewing research articles for renowned journals/conferences and has also participated as a session chair in different conferences. Her research interests include security aspects of multimedia (image, audio, and video), compression, visual privacy, encryption, steganography, secure transmission in future networks, video quality metrics, key management schemes, computer vision algorithms, deep learning models, and General Data Protection Regulation (EU-GDPR).

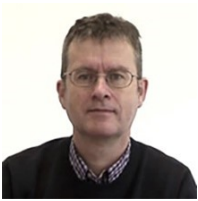

Brian Lee is the Director of the Software Research Institute, Athlone Institute of Technology, Athlone, Ireland. He received the Ph.D. degree from Trinity College Dublin, Dublin, Ireland, in the application of programmable networking for network management. He has over 25 years R&D experience in telecommunications network monitoring, their systems and software design and development for large telecommunications products with very high impact research publications. Formerly, he was the Director of research for LM Ericsson, Ireland, with responsibility for overseeing all research activities, including external collaborations and relationship management. He was an Engineering Manager with Duolog Ltd., where he was responsible for strategic and operational management of all research and development activities.